\documentclass[a4paper,pre,10pt,amsmath,amssymb,nofootinbib,showpacs,superscriptaddress,floatfix,twocolumn]{revtex4}
\usepackage[english]{babel}
\usepackage{graphicx}

\newcommand{\dfr}[2]{\frac {\displaystyle #1}{\displaystyle #2}}
\usepackage[usenames]{color}

\begin{document}
\title{Quantum--classical crossover close to quantum critical point}
\author{Mikhail Vasin}
\address{Physical-Technical Institute, Ural Branch of Russian Academy of Sciences, 426000 Izhevsk, Russia}
\address{High Pressure Physics Institute, Russian Academy of Sciences, Moscow, Russia}
\author{Valentin Ryzhov}
\address{High Pressure Physics Institute, Russian Academy of Sciences, Moscow, Russia}

\begin{abstract}
We analyze the quantum-classical crossover in the vicinity of the
continuous quantum critical point (QCP) of a Boson system. The
analysis is based on the Keldysh approach for the description of
of the non-equilibrium quantum dynamics. The critical behavior
close to QCP has three different regimes (modes): adiabatic
quantum mode (AQM), dissipative classical mode (classical critical
dynamics mode (CCDM) and dissipative quantum critical mode (QCDM).
Crossover among these regimes (modes) is possible:  it is shown
that the experimentally observed changing of the critical
exponents close to QCP is the dynamical effect accompanying the
crossover from CCDM where the thermal fluctuations dominate to
QCDM where the quantum fluctuations determine the critical
behavior. In this case the effective dimension of the $d$-dimensional system
continuously changes from $D_{eff}=d$ to $D_{eff}=d+2$ while the
universality class of the system does not change.
\end{abstract}

\maketitle

Recently there has been considerable interest in the experimental
and theoretical studies of the quantum phase transition dynamics.
This interest is quite natural as quantum phase transitions are
essentially dynamic \cite{Hertz}. In this case the time acts as an
additional space dimension
\cite{Sachdev,QT1,QT2,QT3,QT4,QT5,QT6,QT7}. However, usually one
considers only a simplified case assuming that in the vicinity of
the critical point it is possible to distinguish two regimes: in
one of them the energy of thermal fluctuations exceeds the energy
of quantum fluctuations, $k_{B}T\gg \hbar \omega_{\Gamma }$
($\omega_{\Gamma }$ is the quantity reciprocal to the relaxation
time of the system, $\tau_{\Gamma}$), the critical mode being
described by classical dynamics; in the other one the energy of
thermal fluctuations becomes less than the energy of quantum
fluctuations, $k_{B}T\ll \hbar \omega_{\Gamma }$, the system being
described by quantum mechanics \cite{Hertz,Sachdev}. This
description is not complete since it does not take into account
the effect of dissipation in the quantum fluctuation regime,
though it is well known that dissipation drastically change the
critical properties \cite{L,W,W1,W2}. It is clear that the system
turns from the mode of dissipative dynamics of thermal
fluctuations into the adiabatic mode of purely quantum
fluctuations, then there should exist some intermediate
dissipative mode of quantum fluctuations. The crossover between
these critical modes has not been theoretically studied so far. It
will be shown below that within a unified approach based on the
Keldysh technique of non-equilibrium dynamics description, the
crossover among all three critical modes in the vicinity of the
quantum critical point will be described. The special attention
will be devoted to the to the experimentally observable situation
\cite{Erkelens} when the crossover from the classical to the
quantum criticality takes place with the corresponding continuous
change of the critical indexes. Below we will show, that in this
case the system universality class does not change. The
modification of the critical indexes is the result of the change
of the effective dimension of the dynamic system, which occurs at
the transition from the thermal fluctuations to the quantum
fluctuations.

To describe quantum critical dynamics theoretically, it is most
convenient to use the Keldysh technique initially formulated for
quantum systems. Let the system of our interest be the Boson
system, whose state is described with the scalar field of the
one-component order parameter $\phi $, and the potential energy is
determined by the functional $U(\phi )$, e.g. $U\propto \phi^4$.
Let us assume that $\hbar =1$ and $k_{B}=1$. In the static, to say
more correctly, in the stationary, not quantum case the physics of
the system is determined by the partition function:
\begin{gather*}
Z=N\int \mathfrak{D}\phi\exp \left[-S(\phi)\right],
\end{gather*}
where $\int \mathfrak{D}\phi$ denotes the functional $\phi $-field
integration, $S$ is the action that in the general form is as
follows:
\begin{gather*}
S(\phi)=\dfr 1{T}\int dk \left( \phi^{\dag} G^{-1}\phi + U(\phi)\right),\\
G^{-1}=\varepsilon_k=k^2+\Delta,
\end{gather*}
where $T$ is temperature of the system, $U=v\phi^4$, $\Delta $ is the
governing parameter, that tends to zero at the critical point.

There are different methods for the description of the transition
from equilibrium statics to non-equilibrium dynamics. Note, that
all of them result in doubling the fields describing the system,
suggest the interaction of the system with the thermostat and are
essentially equivalent to each other. As it has been mentioned
above, we are going to use the Keldysh technique, which seems most
convenient. In this case the role of the partition function is
played by the functional path integral that after Wick rotation
has the form \cite{Kamenev}:
\begin{gather*}
\displaystyle Z=N\int \mathfrak{D}\phi^{cl} \mathfrak{D}\phi^{q}\exp \left[-S(\phi^{cl}, \phi^{q})\right],\\
\displaystyle S(\phi^{cl}, \phi^{q})=\int d\omega dk \left( \bar\phi^{\dag} \hat G^{-1}\bar\phi + U(\phi^{cl}+\phi^q)\right. \\
\displaystyle \left.-U(\phi^{cl}-\phi^q)\right),\quad \bar\phi =\{\phi^{q},\,\phi^{cl}\},
\end{gather*}
where $\phi^{q}$ and $\phi^{cl}$ are pair of fields called
``quantum'' and ``classical'' respectively. In the case of the
Boson system the matrix of the inverse correlation functions is
the following \cite{Kamenev}:
\begin{gather}
\hat G^{-1}=\left[ \begin{array}{cc}\displaystyle 0 & \omega^2+\varepsilon_k+i\Gamma\omega \\
\displaystyle \omega^2+\varepsilon_k-i\Gamma\omega & 2\Gamma \omega \coth(\omega/T) \end{array}\right],
\label{R1}
\end{gather}
where $\Gamma $ is the kinetic coefficient, and the function
$f(\omega,\,T)=\coth(\omega/T)$ is the function of the density of
states of the ideal Boson gas. The advancing, retard and Keldysh
parts of both the correlation functions matrix and the inverse
matrix are connected by the relation known as the
fluctuation-dissipation theorem (FDT): $\displaystyle [\hat
G^{-1}]^{K}=2\coth(\omega/T)\,\mbox{Im} \left([\hat
G^{-1}]^{A}\right)$.

The expressions given above allow us to describe the critical
dynamics of the system theoretically in the vicinity of the
critical point. They are general and allow the system to be
described both within the classical, $T\gg \omega$,  and the
quantum, $\omega \gg T$, limits.

Let us consider the first region, where thermal fluctuations
dominate, $\omega \ll T$. Note that the plane $\omega =0$ is
entirely located in this region. The critical dynamics of the
system is determined by the Keldysh element of the matrix of Green
functions, $[\hat G^{-1}]^K=2\Gamma \omega \coth(\omega/T)$.
Within $T\gg \omega$ this function tends to $\lim\limits_{T\gg
\omega } [\hat G^{-1}]^K \approx 2\Gamma T$. Note that in this
case the effect of the thermostat on the system (the action of the
statistical ensemble on its own element) corresponds to the
influence of the external ``white'' noise. The fluctuations with
the smallest wave vectors and energies ($k\to 0$, $\omega\to 0$)
are considered to be relevant (significant) in the vicinity of the
critical point, hence only the terms with the lowest degrees $k$
and $\omega $ are retained in the expressions. As a result, in the
fluctuation field the system is described dy the standard
classical non-equilibrium dynamics:
\begin{gather*}
\hat G^{-1}=\left[ \begin{array}{cc}0 & \varepsilon_k+i\gamma\omega \\
\varepsilon_k-i\Gamma\omega & 2\Gamma T \end{array}\right].
\end{gather*}
satisfying the standard form of FDT:
\begin{gather*}
[\hat G^{-1}]^{K}=({T}/{\omega })\mbox{Im} \left([\hat G^{-1}]^{A}\right).
\end{gather*}
The dispersion relation in this case is: $\omega \propto k^2$,
whence it follows that the dynamic critical exponent of the theory
(scaling dimension) in the first approximation will be: $z=2$. The
dimension of the system is: $D=d+z=d+2$, but due to the presence
of``white'' noise  the effective dimension of the system is:
$D_{eff}=D-2=d$ \cite{Parisi}. Naturally, it results in the
coincidence of the critical dimensions of the dynamic and static
theories, the critical behavior of the system being described by
the classical critical dynamics of the  $d$-dimensional system.
Let us refer to this mode as the mode of the classical critical
dynamics (CCDM) ($T\gg \omega, \Gamma T\gg |\Delta|$).

Now let us consider the case when the quantum fluctuations
dominate, $\omega\gg T$. In this case (\ref{R1}) has the form:
\begin{gather*}
\hat G^{-1}\approx \left[ \begin{array}{cc}0 & \varepsilon_k+i\Gamma\omega \\
\varepsilon_k-i\Gamma\omega & 2\Gamma |\omega | \end{array}\right],
\end{gather*}
so the system ``does not know'' that it has got temperature.
In spite of the absence of thermal fluctuation in the quantum case
FDT still exists and has the following form:
\begin{gather*}
[\hat G^{-1}]^{K}=2\,\mbox{sign}(\omega )\mbox{Im} \left([\hat G^{-1}]^{A}\right),
\end{gather*}
and the action of the statistic ensemble on the system does not
depend on the temperature. Note, that close to the phase
transition ($\Delta \approx 0$), when $\varepsilon _k\to 0$, we
get $G^K(\omega )={2}/(\Gamma |\omega |)$. This is the so called
$1/f$-noise (Flicker noise), whose intensity does not depend on
the temperature. The latter significantly changes the critical properties of the system.  As in the case of classical critical dynamics the dimension in this case is $D=d+2$. However, the $1/f$ noise in contrast to the ``white''-noise, does not decrease the effective dimension  \cite{Vasin}, therefore the effective dimension of the dissipative quantum system is greater by 2 than its static dimension, $D_{eff}=d+2$. The disagreement of the static and dynamic theories is accounted for by the fact that in the quantum case there is no statistic limit, and the only correct results are those of the dynamic theory. This dynamic mode can be referred to as the mode of the quantum critical dynamics (QCDM) ($T\ll \omega, \Gamma \omega\gg |\Delta|$).

With $\omega \gg T$ also the case is possible when
the coherence time of the system appears
much shorter than the inverse frequency of quantum fluctuations,
$\Gamma \omega  \ll |\Delta |$, the dynamics of the system changes
into an adiabatic mode, in which the dissipation can be neglected, $\Gamma \to 0$, thus:
\begin{gather}\label{S1}
\hat G^{-1}\approx \left[ \begin{array}{cc}0 & \omega^2+\varepsilon_k\\
\omega^2+\varepsilon_k & 0\end{array}\right],
\end{gather}
the dispersion relation takes the form $\omega \propto k$, and as
a result, $z=1$. In this region the critical behavior is described
as the critical behavior of the static system with the dimension
$D_{eff}=d+1$. It is easy to see that
this is the region in which the Matsubara formalism works.
Also one can see that in this case the critical behavior of a
three-dimension system has the simple description within the mean
field theory (Ginzburg--Landau theory), since the effective dimension is equal to the critical dimension, $D_{eff}=d_c^+=4$.
This regime can be referred to the adiabatic quantum mechanical mode (AQM) ($T\ll \omega, \Gamma \omega\ll |\Delta|$).

\begin{figure}[h]
\centering
   \includegraphics[width=7.5cm]{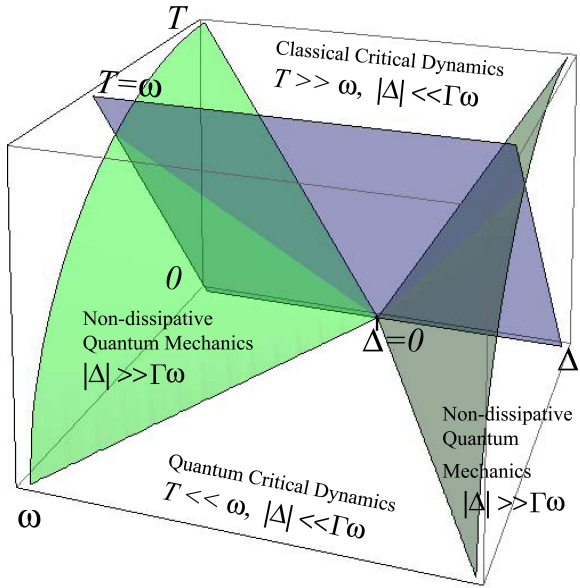}
   \caption{The green colour denotes the conventional surface $\omega^2+T^2=|\Delta|^{2\nu}$ showing
   the location of the crossover region between the dissipative and adiabatic fluctuation modes.}
   \label{F2}
\end{figure}

\begin{figure}[h]
\centering
   \includegraphics[width=8cm]{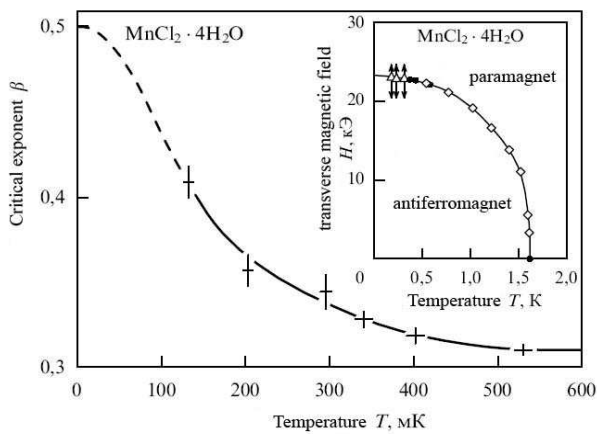}
   \caption{The dependence of critical exponent $\beta $ determining the order parameter behavior,
   on the temperature of the antiferromagnet MnCl$_2\cdot $4H$_2$O\cite{Erkelens}. }
   \label{f4}
\end{figure}

The schematic picture of the different critical regions is shown
in Fig.\,\ref{F2}, the surfaces indicate the regions of the
crossover between the critical modes. Let us consider separately
the region of ``crossing'' of all modes, which is in the vicinity
of the plane $\omega =T$. Here the thermal and quantum
fluctuations are equal, thus this area is the region of crossover
between classical and quantum dynamic modes. The crossover from
CCDM to QCDM can be observed experimentally \cite{Erkelens}. For
example, the experimental dependence of critical index $\beta $ on
the temperature is shown in Fig.\,\ref{f4} \cite{Erkelens}. At the
relatively high temperatures, $T/\omega\gg 1$, this exponent
correspond to the value characteristic for the three-dimensional
classic system. However at small temperatures, $T/\omega\ll 1$, it
becomes equivalent to the corresponding critical exponent of mean
field theory, since the effective dimension of the system becomes
greater than the critical dimension: $D_{eff}=d+z\geqslant d_c^+$.

At first sight this dependence of the critical exponents on the
temperature seems strange, because the critical exponents are
dependent on the system universality class, which can not
continuously change with temperature. However, in the critical
dynamics the critical exponents do not only depend on the
universality class, but also on the nature of the critical
fluctuations. Therefore, the reason of this continuous change of
the critical exponents is the crossover from the thermal
fluctuations mode to the quantum fluctuations mode with the
corresponding increase of the effective dimension, while the
system's universality class does not change.

One can show that this crossover is driven by the temperature
dependence of the density of states.  The point is that all
diagrams giving the main contribution in the renormalization
theory, contain the loop consisting of one retarded (or advanced)
Green function, and one Keldysh Green function \cite{Vasin}.  In
the long-wavelength limit these contributions have the form:
\begin{gather}
\sim \int {2\Gamma \omega f(\omega,\,T)}/{\omega ^3}d\omega ^{1+d/2},
\label{A0}
\end{gather}
where $\Gamma $ is dimensionless. Therefore the integral in
(\ref{A0}) logarithmically diverges when $\omega f(\omega,\,T)\sim
\omega ^{2-d/2}$. There is no longer any difference from the usual
renormalization procedure. If the function $\omega f(\omega )$
would be the power function, $\omega f(\omega )=\omega^{\Lambda
}$, then the problem is reduced to the usual problem of the
investigation of the critical behavior of $d+2\Lambda
$-dimensional system.

Unfortunately the function $\omega f(\omega,\,T)=\omega
\coth(\omega/T)$ is the complicated non power one. This fact
seemingly does useless our usual reasoning for deriving of the
temperature dependence of the critical exponents in analytical
form. However, the slope tangent to this function continuously changes in the bounded limits: from 0 at $\omega/T\ll 1$  to 1 at $\omega/T\gg 1$ \cite{Vasin}. Since in
the experiment $\omega/T$ is controlled by the temperature and the
characteristic energy of the quantum fluctuations, $\omega_0$, then for
every value $\omega_0/T$ the function of density of states can be
approximated by the power function $\omega f(\omega,\,T)\approx
\omega ^{\Lambda(\omega_0/T)}$, where $\Lambda(x)=x\partial \ln [x
f(x)]/\partial x$ ($0\leqslant \Lambda \leqslant 1$), and
$\omega_0$ is the characteristic frequency of the system, which
depends the quantum fluctuation energy. Thus, the exponent
$\Lambda $ is the only function of temperature
$\Lambda(\omega_0/T) = (\coth(\omega_0/T) -
x\,\mbox{csch}^2(\omega_0/T)) \tanh(\omega_0/T)$.

The above approximation allows us to use the conventional
renormalization procedure for the calculation of the critical
indexes. If at the temperature $T'$ the function of density of
states can be approximated by some power function $\omega
f(\omega,\,T')\sim \omega ^{\Lambda (T')}$, then the critical
behavior of the system is identical to the classical (non-quantum)
critical behavior of the $(d+2\Lambda )$-dimensional system.

In order to estimate the temperature dependence of the critical
exponent $\beta $ we can use the well known relations for the
critical exponents: $\alpha +2\beta+\gamma=2,\quad d\cdot \nu
=2-\alpha ,\quad \gamma =\nu\cdot(2-\eta )$. These exponents
characterize the heat capacity, $C_v\sim |\Delta |^{-\alpha }$,
susceptibility, $\chi\sim |\Delta |^{-\gamma }$, magnetization,
$\langle \phi \rangle \sim |\Delta |^{\beta }$, correlation
radius, $r_c\sim |\Delta |^{-\nu }$, and Green function, $G(r)\sim
r^{-d+2-\eta }$ ($\eta $ is the anomalous dimension index).
According to above, in the case of the crossover from CCDM to QCDM, when $d\to d'=d+2\Lambda
$, relations for the critical exponents are valid for efficient
exponents: $\nu\to \nu '(\Lambda )$, $\eta \to \eta '(\Lambda )$, $\beta \to \beta '(\Lambda )$,
$\gamma \to \gamma '(\Lambda )$,  $\alpha \to \alpha '(\Lambda )$.

\begin{figure}[h]
\centering
   \includegraphics[width=8cm]{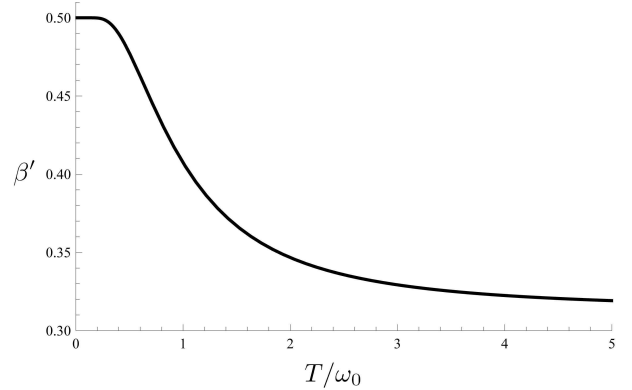}
   \caption{Theoretical dependence of critical exponent $\beta '$ on the $T/\omega_0$ ratio for the $\phi^4$-model.}
   \label{f6}
\end{figure}

First dependence can be derived from the well known
renormalization group equations (in the one-loop approximation) \cite{Doc}:
\begin{gather*}
\dfr{d\ln |\Delta |}{d\xi }=2-v\dfr 3{8\pi^2}, \quad \dfr{d\ln v}{d\xi} =\varepsilon' -v\dfr 9{8\pi^2},
\end{gather*}
where $\xi $ is the regularization parameter, $\varepsilon'
=(4-d')/2=(4-d)/2-\Lambda=\varepsilon -\Lambda$. From the fixing
condition for $v$, $dv/d\xi=0$, and definition $\nu'\equiv (d\ln
|\Delta |/d\xi)^{-1}$ one can get:
$\nu'=1/(2-\varepsilon'/3)=1/(\nu^{-1}+(2-\nu^{-1})\Lambda)$.
In the three-dimension case $\varepsilon =1$, and $\nu\approx 0.642$ \cite{Doc}.
Because $\eta \approx 0$ for all dimensions,
then we
have $\gamma'\approx 2\nu'$, and $(d+2\Lambda )\cdot
\nu'=2(\beta'+\nu')$. As a result
$\beta'=(d/2+\Lambda-1)\nu'$. Using these
formulas and the dependence $\Lambda(\omega_0/T)$ we can calculate
functions $\beta' (T)$, $\nu '(T)$, $\gamma' (T)$, and $\alpha' (T)$ (Fig.\,\ref{f6}).  One can see that $\beta' (T)$ dependence is in good qualitative agreement with experimental data
\cite{Erkelens}.

From above one can see that the critical behavior in the vicinity
of the quantum critical point is multi-critical. The functional
technique of theoretical description of non-equilibrium dynamics
allows us to describe the entire spectrum of critical modes in the
vicinity of quantum phase transition within a single formalism.
In particular, it describes the crossover between CCDM and QCDM,
and the unusual temperature dependence of the system critical
exponents. Note that in this case the system universality class
does not change. The continuous change of the critical exponents
is the dynamical effect, which is caused by the crossover from the
thermal fluctuation mode to the quantum fluctuation mode.

We are grateful to S. M. Stishov and  V. V. Brazhkin for
stimulating discussions. This work was partly supported by the
RFBR grants No. 13-02-91177 and No. 13-02-00579.


\begin{thebibliography}{99}

\bibitem{Hertz} J. A. Hertz, Phys. Rev. B {\bf 14,} 1165 (1976).

\bibitem{Sachdev} S. Sachdev, {\it Quantum Phase Transitions, Cambridge University Press, New York},
ISBN 978-0-521-51468-2, 501 p., 2011.

\bibitem{QT1} S. L. Sondhi, S. M. Girvin, J. P. Carini, and D.
Shahar, Rev. Mod. Phys. {\bf 69}, 315 (1997).

\bibitem{QT2} Matthias Vojta, Rep. Prog. Phys. {\bf 66},  2069 (2003).

\bibitem{QT3} S. M. Stishov, Phys. Usp. {\bf 47}, 789 (2004) (DOI:
10.1070/PU2004v047n08ABEH001850).

\bibitem{QT4} S. Sachdev, Nature Physics {\bf 4}, 173 (2008).

\bibitem{QT5} P. Gegewart, Q.Si, and F. Steglich, Nature Physics {\bf 4}, 186 (2008).

\bibitem{QT6} T. Giamarchi, C. Ruegg, and O. Tchernyshov, Nature Physics {\bf 4}, 198 (2008).

\bibitem{QT7} D. M. Broun, Nature Physics {\bf 4}, 170 (2008).



\bibitem{L} A.J. Leggett, S. Chakravarty, A.T. Dorsey, M.P.A. Fisher, A. Garg, and W.
Zwerger, Rev. Mod. Phys. {\bf 59,} 1 (1987).

\bibitem{W} U. Weiss, {\it Quantum Dissipative Systems. World Scientific, Singapore,
1999}.

\bibitem{W1} P. Werner, K. V\"{e}olker, M. Troyer, and S. Chakravarty, Phys. Rev. Lett. {\bf 94,} 047201 (2005).

\bibitem{W2} P. Werner, M. Troyer, and S. Sachdev, J. Phys. Soc. Jpn. Suppl. {\bf 74,} 67 (2005).

\bibitem{Erkelens} W.A. Erkelens {\it et al}. Europhys. Lett. {\bf 1,} 37 (1986).

\bibitem{Kamenev} Alex Kamenev, {\it Field theory of non-equilibrium systems. Cambridge University Press, New York, 2011} ISBN 978-0-521-76082-9.

\bibitem{Parisi} G. Parisi, N. Sourlas, Phys. Rev. Lett. {\bf 43,} 744 (1979).

\bibitem{Vasin} M.G. Vasin, Physica A {\bf 415,} 533 (2014).

\bibitem{Doc} V.S. Dotsenko, Phys. Usp. {\bf 38,} 457 (1995).



\end{thebibliography}
\end{document}